# SINGULARITIES OF MAGNETIC AND ELASTIC CHARACTERISTICS OF La$_{2/3}$Ba$_{1/3}$MnO$_3$: ANALYSIS OF MARTENSITIC KINETICS

E.L. Fertman[a,*], A.B. Beznosov[a], V.A. Desnenko[a], L.N. Pal-Val[a], P.P. Pal-Val[a] and D.D. Khalyavin[b]

[a]B. Verkin Institute for Low Temperature Physics and Engineering NASU,
47 Lenin Ave., Kharkov 61103, Ukraine

[b]Institute of Solid State and Semiconductor Physics NASB, 17 P. Brovka str.,
Minsk 220072, Belarus

**Abstract**

A coordinated temperature behavior of magnetic susceptibility and internal friction has been observed in the La$_{2/3}$Ba$_{1/3}$MnO$_3$ manganite in the temperature region of the crystal phase separation 5-340 K. Stepwise temperature behavior of the susceptibility of the single crystal sample and corresponding singular behavior of the internal friction in the polycrystalline manganite have been found. These small-scale features of the temperature dependences of the susceptibility and the internal friction are considered to be a reflection of martensitic kinetics of the structural phase transformation $R\bar{3}c \leftrightarrow Imma$ in the 200 K temperature region.

* Corresponding author. Tel.: +38-057-3308503; fax: +38-057-3450593.
*E-mail address*: fertman@ilt.kharkov.ua

## 1. Introduction

$La_{2/3}Ba_{1/3}MnO_3$ manganite is one of the compounds in which the colossal magnetoresistance (CMR) effect has been found [1, 2, 3, 4]. Besides, the compound is known to reveal interesting phenomena accompanying the structural phase transformation of martensitic type in 200 K temperature region due to the strong interplay between electronic structure, lattice and magnetic properties [5, 6].

Martensitic transformations are first order solid-solid diffusionless structural phase transformations. Unlike other phase transformations, where diffusion disperses the neighboring atoms, here neighboring atoms in the parent phase also remain in the product phase [7]. However, the lattice gets distorted due to spontaneous displacement of the atoms (from their positions in the parent lattice) accompanying the discrete change in the shape and symmetry of the unit cell. This creates long-range strain fields, which in turn strongly depend on the relative positions and orientations of the martensitic plates. The transformation path depends on the continuously evolving strain fields that leave the system in a metastable state.

Structural phase segregation in manganites with reference to the martensite transition is considered mostly in the insulating narrow-bandwidth compounds experiencing charge ordering. Mesoscopic (500-2000 Å) phase segregation in $Pr_{0.73}Ba_{0.3}MnO_3$ was studied by neutron scattering in Refs. [8,9]. Characteristic martensitic behavior was observed in $Bi_{0.2}Ca_{0.8}MnO_3$, $Pr_{0.5}Ca_{0.5}MnO_3$ and $La_{0.275}Pr_{0.35}Ca_{0.375}MnO_3$ by polarized optical microscopy below 180, 230 and 210 K, respectively [10]. Magnetization steps in the field as well as time dependences in $Pr_{0.5}Ca_{0.5}Mn_{0.97}Ga_{0.03}O_3$ and $Pr_{0.5}Ca_{0.5}Mn_{0.95}Ga_{0.05}O_3$ were observed in Refs. [11,12] and considered as a manifestation of the martensitic character of the processes. Analogous phenomena were studied in the thick (>0.5 μm) films of $Pr_{0.65}(Ca_{0.8}Sr_{0.2})MnO_3$ in Ref. [13]. Magnetoresistive memory, as a feature of the martensitic behavior of $La_{0.325}Pr_{0.300}Ca_{0.375}MnO_3$, was studied in Ref. [14]. Authors of Ref. [15] studied structural phase separation below 180 K in $La_{1-x}Ba_xMnO_3$ ($x\sim0.5$) with charge ordering in the A- and B- sites.

Phenomena resembled martensitic transitions were observed in the intermediate- and large-bandwidth manganites with rather high conductivity below the Curie point as well. It is evident, that nature of these transitions is different from one of transitions connected with charge ordering. Coexistence of the *Pbnm* and $R\bar{3}c$ phases in La$_{0.8}$Ba$_{0.2}$MnO$_3$ within the interval 185<T<196 K was studied in Refs. [16,17]. The samples represented single crystals, the first order orthorhombic↔rhombohedral phase transition occurred in the ferromagnetic state (below the Curie point T$_c$≈255 K) at T$_s$≈175 K. Pronounced temperature hysteresises of magnetic susceptibility and electrical resitivity associated with this transition were revealed. Extended temperature hysteresis of elastic modules and resitivity was also found in the La$_{1-x}$Sr$_x$MnO$_3$ single crystals with $x$ between 0.15 and 0.2 [18,19].

In principal the natural media for martensitic transitions are single crystalline systems. In polycrystals with small enough grains growth of martensitic domains across grain boundaries is prohibited, the same is valid for thin films and small particles as well. Another reason is that the surface energy affects substantially the competition of various crystal structures and may prevent their exchange in thin films and small particles. This view is being confirmed by experiments: as one can see from resistivity measurements, a temperature anomaly manifested structural phase transition in the bulk La$_{2/3}$Ba$_{1/3}$MnO$_3$ [20] is absent in the 150 nm film grown on SrTiO$_3$ substrate [21]; for the La$_{0.95}$Ba$_{0.05}$MnO$_3$ films and La$_{2/3}$Ba$_{1/3}$MnO$_3$ superlattices the dependence of properties on the thickness and type of substrate (SrTiO$_3$ and NdGaO$_3$) has been demonstrated in Refs. [22] and [23], respectively, and no sign of the 1$^{st}$ order structural transition was observed.

The x-ray and neutron diffraction study [3] has shown, that La$_{2/3}$Ba$_{1/3}$MnO$_3$ manganite possesses the rhombohedral crystal structure of the $R\bar{3}c$ space group at room temperature, and orthorhombic structure of the *Imma* space group at 4 K. Following the group theory the structural transformation $R\bar{3}c$↔*Imma* has to be of the 1$^{st}$ type only [24]. An important detail of this transition is sudden rearrangement of tiltings of the MnO$_6$ octahedra [25]. Authors of the Refs. [5,6] observed process of exchange of the crystallographic phases in La$_{2/3}$Ba$_{1/3}$MnO$_3$ in the wide temperature interval, and noted, that the phenomenon resembled a martensitic transformation. But it would be advisable to confirm the supposition by additional experiments. Along with an extended temperature region of the transformation another characteristic feature of martensitic transition is a burstlike growths of the martensitic phase domains (in essence, this phenomenon is the central feature of the martensitic kinetics and thus, of the transformation as a whole). Signs of such a process could be detected in elastic and, perhaps, magnetic characteristics when temperature scanning, so their detailed study would be of interest. The present study has revealed some features of kinetics of the phase transformation studied, reflected in a singular character of magnetic susceptibility and internal friction temperature dependences.

## 2. Experiment

Experiments were carried out with single crystalline and polycrystalline samples prepared in the Institute of Solid State and Semiconductor Physics NASB. The single crystal of $La_{2/3}Ba_{1/3}MnO_3$ composition used for the magnetic measurements only was prepared from $B_2O_3$-$BaO$-$BaF_2$ flux by slow cooling in air from $1150°C$ in a platinum crucible [26]. Other samples used both for the ultrasonic and magnetic experiments were made from polycrystalline (ceramic) billets of the $La_{2/3}Ba_{1/3}MnO_3$ compound which were prepared using standard solid-state reaction with stoichiometric amounts of powders $La_2O_3$, $BaCO_3$, and $Mn_2O_3$; the details are similar to those published in Ref. [5]. Samples quality was confirmed by x-ray diffraction study. The grain size in the crystallites amounts 10-15 μm. Microstructure is expected to be a twin-type lamella one, the twin crystallization occurs at about 548 K [27].

A.c. magnetic susceptibility was measured by the induction technique in magnetic fields of the $H_0$=2.4 Oe amplitude and frequencies $10^2$-$10^4$ Hz between 5 K and 370 K. The samples size was 2.7×2.5×2.0 $mm^3$, and the demagnetization factor, respectively, was $N'$=3.6 in the long side direction, along which the external magnetic field was applied. Taking into account the value of magnetic susceptibility of $La_{2/3}Ba_{1/3}MnO_3$ $\chi$~10÷14 [5], and relation $H_i$=$H/(1+N\chi)$ between external H and internal $H_i$ magnetic fields in the sample [28], one can suppose, that amplitude of the effective magnetic field in the samples was $H_{0i}$≤$10^{-2}$ Oe.

Temperature dependence of ultrasound characteristics was obtained by the two-component oscillator technique for longitudinal standing waves [29] at frequencies around 70 kHz between 5 K and 340 K. The sound wave vector was oriented along the long side of the 25×4.44×2.65 $mm^3$ size sample. Compressibility κ and internal friction $Q^{-1}$ of the compound were determined in accordance with Ref. [29].

Both at the magnetic as well as ultrasonic experiments an average time of the temperature points stabilization during the measurement process was 3.6 min; average rate of the temperature changes in the 5 K-340 K interval was 1.2 K/min.

## 3. Results and discussion

*3.1. Thermal hysteresis*

Temperature dependences of the magnetic and ultrasound characteristics obtained well correspond to previous data on neutron diffraction and magnetometry concerning the character of the phase transformation in the temperature range 140-370 K. According to Ref. [5, 6] below the room temperature crystal structure of $La_{2/3}Ba_{1/3}MnO_3$ manganite represents a mixture of the high temperature rhombohedral $R\bar{3}c$ phase (austenite) and the low temperature orthorhombic phase *Imma* (martensite) as a result of martensitic phase transformation. It occurs over a wide range of temperatures [5, 6] that is typical for transformations of such type [7,30]. Note, that on cooling, a martensitic transformation starts at the martensite start temperature $M_s$, and finishes at the martensite finish temperature $M_f$. In the reverse heating cycle, the transformation starts at the austenite start temperature $A_s$ and finishes at the austenite finish temperature $A_f$, which is usually higher than $M_s$. Thus we have certain irreversibility of the structural domain wall motion on cooling and heating, which typically leads to the thermal hysteresis of physical properties of martensite compounds. As it was expected, the pronounced hysteretic behavior of the a.c. susceptibility $\chi_{ac}(T)$ (Fig. 1 for the single crystal and Fig.2 for the ceramic sample) and elastic characteristics (Fig. 3 for the internal friction $Q^{-1}$ and Fig.4 for the reduced compressibility $\kappa(T)/\kappa(340)$) of $La_{2/3}Ba_{1/3}MnO_3$ compound has been revealed. The susceptibility hysteresis loop is asymmetric: its high temperature extent (~140 K) is much greater than the low temperature one (~20 K). This is in accordance with the neutron diffraction data [5, 6] and is attributed to the high temperature of the reverse martensitic transformation that finishes at ~370 K when heating (strictly speaking the $A_f$ temperature for the present samples can be slightly different from 370 K because of certain difference between their oxygen content and that of the samples have been studied in [5,6]; see next Section).

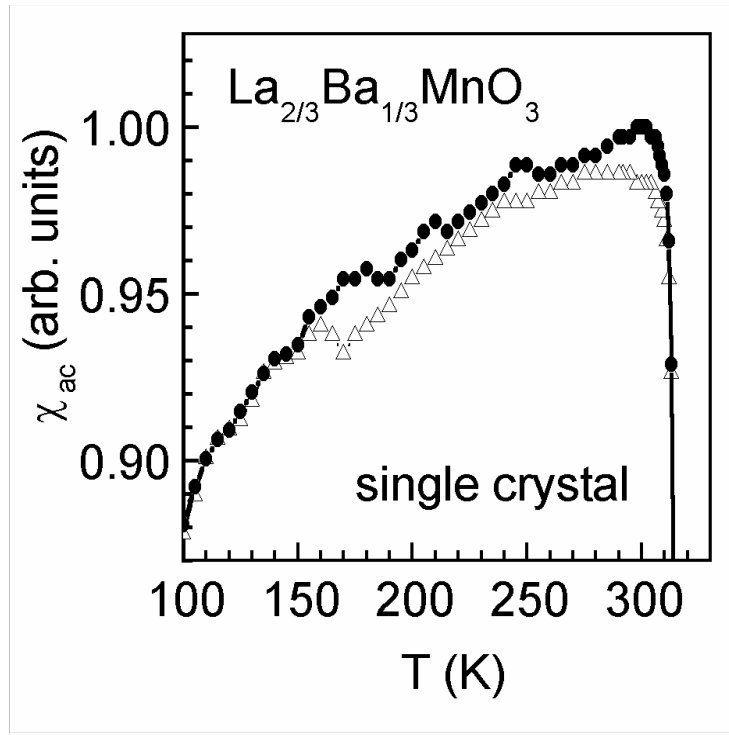

Fig. 1. Temperature dependence of the a.c. magnetic susceptibility at frequency 1 kHz for the $La_{2/3}Ba_{1/3}MnO_3$ single crystal when cooling ($\Delta$) and heating ($\bullet$). The size of the symbols corresponds to the maximal error. Lines are guides for eyes.

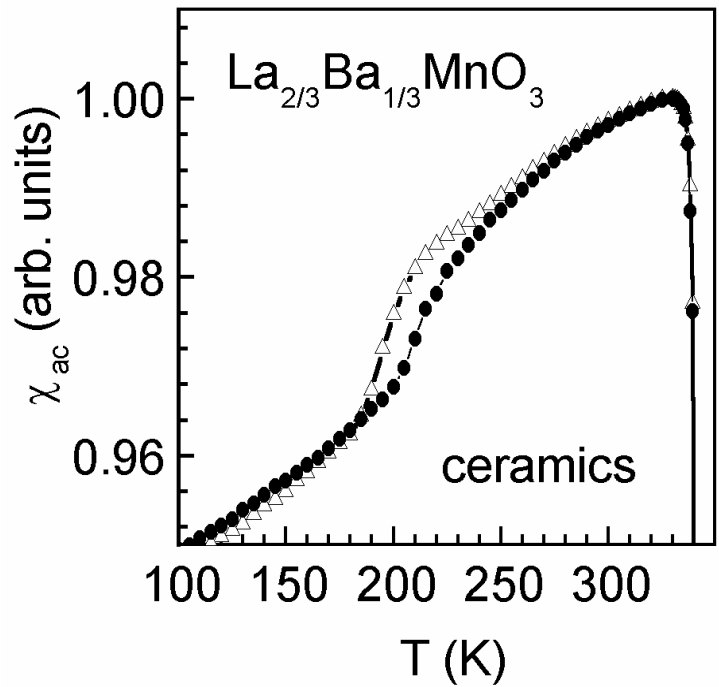

Fig. 2. Temperature dependence of the a.c. magnetic susceptibility at frequency 1 kHz for the $La_{2/3}Ba_{1/3}MnO_3$ ceramic sample when cooling ($\Delta$) and heating ($\bullet$). The size of the symbols corresponds to the maximal error. Lines are guides for eyes.

*3.2. Magnetic structure*

The Curie temperature of the compound $T_C$=340 K appeared to be practically equal for the single crystalline sample and for the ceramic one. This is somewhat higher than it was found for the samples in Ref. [5], and most probably is caused by the higher oxygen index of the present compound. Following to Ref. [2] such a value of the Curie temperature corresponds to the oxygen index 2.99 instead of 2.97, corresponding to $T_C$=314 K in Ref. [5] (note, that as it follows from [3,31] samples with $T_C$~340 K experience the same structural transformation as ones with $T_C$=314 K). This difference between the magnetic ordering temperatures denotes difference of indirect coupling forces between local spins of the manganese ions in the compound. Difference in the features of the interactions in the magnetic system of the compound causes evidently the difference between the shapes of the $\chi(T)$ curves in the present paper and Ref. [5]. Unusual shape of the $\chi(T)$ curve between 250 K and 300 K in [5] was considered as an evidence of presence of a long period magnetic structure (probably a sort of small-angle helix) in the compound. But rather high value $\chi$~14 of the magnetic susceptibility in the temperature range of the supposed helix in [5] implies its low stability [32]. Hence, one can suppose, that the higher temperature of the magnetic ordering in the present paper indicates an increasing of ferromagnetic interaction relatively to antiferromagnetic one in the compound, that makes the above mentioned small-angle helix structure unstable [32], so a ferromagnetic cluster structure appears instead of it. Thus the decrease of the susceptibility with temperature decrease in Figs. 1-2 may be attributed to the increase of the local magnetic anisotropy fields in the system (see e.g. [33]), as it is required by the equation for the initial magnetic susceptibility [28]

$$\chi_0(T) = \frac{M_s^2(T)}{4K(T)},$$

which is evidently measured in the present work. The anisotropy parameter $K(T)$ of the system contains an exchange as well as relativistic contributions [28, 34], and may be composed from the two components, magnetocrystalline and magnetoelastic ones: $K=K_{cr}+K_{me}$. The magnetocrystalline component $K_{cr}$ is dependent on crystal structures, and the magnetoelastic one $K_{me}=K_{as}+K_{gb}$ is induced by mechanical stresses in the system. The $K_{as}$ item is determined by the accommodation stress, accompanying the martensitic transformation, and the $K_{gb}$ item (nonzero for the ceramic sample only) is determined by the stresses induced by grain boundaries due to the anisotropy of the thermal expansion of the crystal grains in ceramics (form of the crystallites changes because of $c/a$ temperature dependence). The lower values of the susceptibility of the single crystal in the cooling process (relatively to their values when heating, Fig. 1) point out a higher $K$ value of the $R\bar{3}c$ phase in comparison with the *Imma* one. If the local magnetoelastic anisotropy substantially exceed the magnetocrystalline one, which is rather low in the nearly cubic crystals, the susceptibility under cooling have to be higher than that under heating, because the elastic stresses in the lattice (and corresponding magnetoelastic contribution to the local anisotropy) increase in the cooling process, and process of their relaxation retards when heating. Such a behavior is connected in any case with the martensitic character of the structural transformation in the studied manganite. We suppose, that the different character of the temperature hysteresis (clockwise one in Fig. 1 and counterclockwise in Fig. 2) is caused by the difference in the $K_{cr}$ and $K_{me}$ contributions to the $K$ values in single crystal and ceramic samples.

*3.3. Steps of the susceptibility*

Another interesting feature revealed is singularity of the temperature behavior of magnetic and ultrasound characteristics of $La_{2/3}Ba_{1/3}MnO_3$ manganite in the phase transformation region (Fig. 1 and Fig. 3), in particular a stepwise shape of the $\chi_{ac}(T)$ curve. On heating the susceptibility curve of the single crystal exhibits bright steps in 150–300 K temperature region which correspond to the reverse martensitic transformation *Imma*→ $R\bar{3}c$ (even between 310 K and $T_C$ the curve had an anomaly). Sharp increase of the susceptibility at 170 K under cooling and decrease at 185 under heating indicate magnetic phase transition (possibly, a sort of spin-reorientation one) in complete agreement with data of Ref. [5] (where size of crystallites was of order of 1 mm, so that the equality $K=K_{cr}+K_{as}$ holds, as it is in single crystals). At the same time the absence of a corresponding anomaly in the ceramic sample may be caused by the suppression of the low field magnetic structure, realized in the single crystal, by higher magnitudes of the effective magnetic anisotropy fields in the ceramic sample (because of high $K_{gb}$ value). Steps of $\chi_{ac}(T)$ above 185 K in the single crystal under heating have evidently another nature (at least because they have no corresponding items under cooling). Sharp falls of the susceptibility against its steady increasing background when heating are consistent with picture of a sharp decrease of volume of domains of the low temperature phase and, correspondingly, of a sharp increase of volume of the high temperature phase domains. When cooling these processes proceed in another manner. Nucleation centers of the *Imma* phase appear in the pure $R\bar{3}c$ phase when cooling uniformly distributed without any coherence between them, so the *Imma* phase fraction growth proceeds by small steps. On the contrary, return (i.e. growth of the residual $R\bar{3}c$ phase clusters) proceeds coherently due to a "memory effect" (the "memory effect" notion is a characteristic item of martensitic ideology [13,14,35,]).Thus the difference in the shapes of the "direct" and "reversal" curves in Fig. 1 and Fig. 3 seems to be natural for martensitic transformations.

*3.4. Ultrasound data*

In the ceramic sample magnetic subsystem is effectively detached from the crystalline anisotropy, so the steps mentioned in the magnetic susceptibility are not seen (Fig. 2). As it was mentioned above, in martensitic transformation growth of martensitic domains across a grain boundary is prohibited. So, grain boundaries are known to affect nucleation and growth of martensitic phases: martensitic properties are seen more brightly when samples are single crystalline [10]. Nevertheless singularities similar to those discussed above for magnetic susceptibilities are seen clearly in the internal friction temperature dependence in Fig. 3. The main characterization of this dependence is well expressed temperature hysteresis in the 150-340 K temperature range and maximal values of $Q^{-1}$ in the 200 K region, where amounts of the both coexisting phases $R\bar{3}c$ and *Imma* are nearly equal (Fig. 3, the insert). On heating certain fine structure of $Q^{-1}(T)$ above its more slowly changing background is formed. It is especially bright in 200 – 250 K temperature region, where the transformation proceeds the most rapidly. The internal friction jumps found imply a sudden growth of amount of the high temperature phase in the compound, reflecting kinetics of the reversal martensitic transformation as well as the stepwise temperature dependence of magnetic susceptibility. As it could be expected, such fine structure has not been found in the compressibility temperature dependence (Fig. 4). Real part of elastic characteristics is less sensitive to the small value dissipation energy in the system, so the only characteristic features of martensitic transformation are the temperature hysteresis of compressibility κ*(T)* in the 150-340 K temperature range, and its maxima in the vicinity of 200 K. This picture is close to those observed in other studies of martesitic transformations (see e.g. [36]).

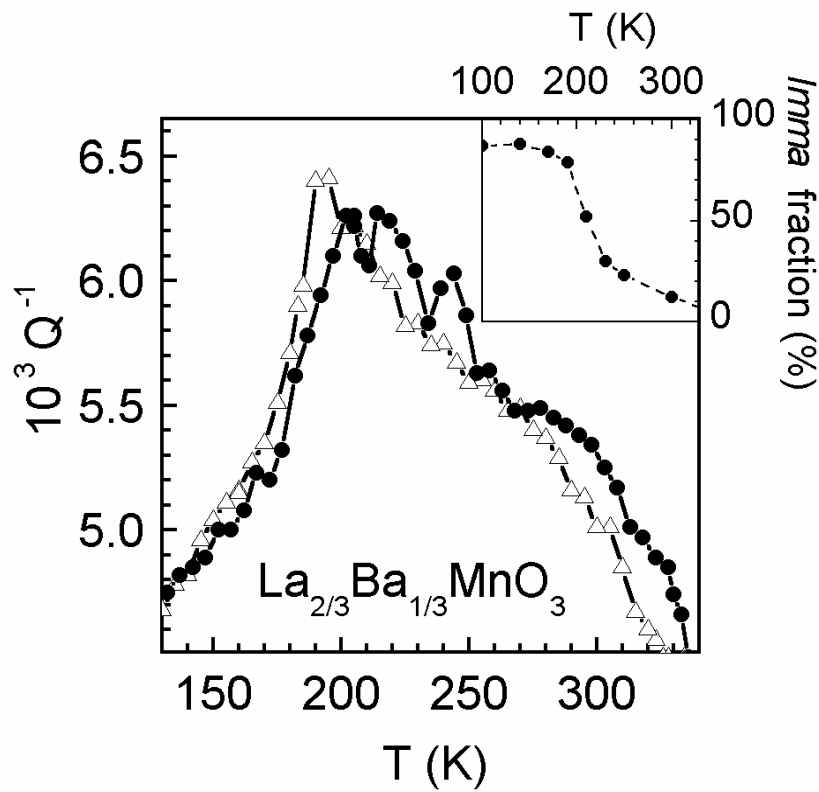

Fig. 3. Temperature dependences of the internal friction $Q^{-1}$ in the ceramic $La_{2/3}Ba_{1/3}MnO_3$ sample when cooling (Δ) and heating (•).The size of the symbols corresponds to the maximal error. The inset shows temperature dependence of the low temperature *Imma* phase fraction [5]. Lines are guides for eyes.

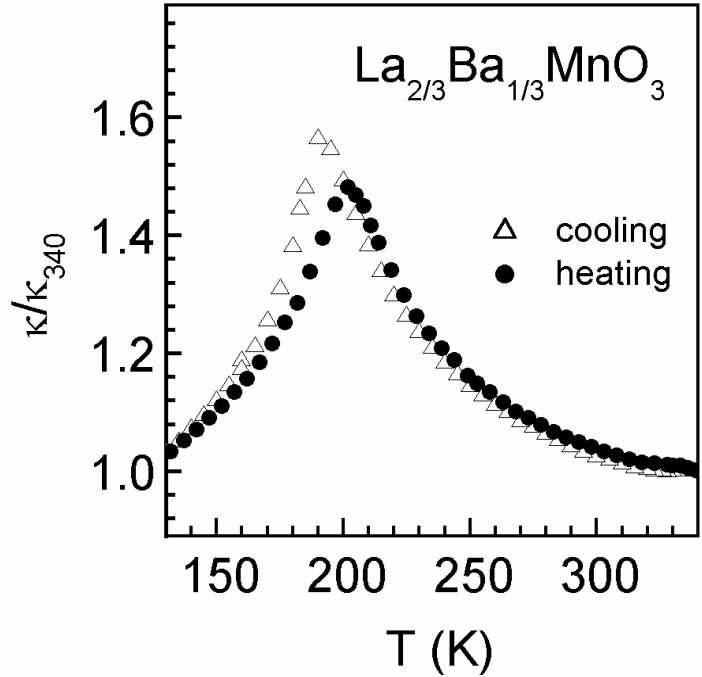

Fig. 4. Temperature dependences of the reduced compressibility $\kappa(T)/\kappa(340)$ of the ceramic $La_{2/3}Ba_{1/3}MnO_3$ sample when cooling (Δ) and heating (•).The size of the symbols corresponds to the maximal error.

## 5. Conclusion

The singular temperature behavior of magnetic susceptibility and internal friction found is consistent with the martensitic scenario of the structural phase separation in the $La_{2/3}Ba_{1/3}MnO_3$ manganite. Sharp stepwise falls of susceptibility and characteristic jumps of internal friction implies to be a result of the discrete process of the phases exchange in the substance, which is characteristic feature of the martensitic transformation kinetics. Temperature intervals of the singularities (from $\Delta T \sim 26$ K in the vicinity of $T \sim 230$ K to $\Delta T \sim 45$ K in the vicinity of $T \sim 280$ K) and magnitudes of the last ones, together with the phase fraction temperature dependence, give information, necessary for an evaluation of the quantity of the discretely evolving martensitic phase, as well of the level of elastic stresses which regulate this process.

The study was supported by grant from the National Academy of Sciences of Ukraine № 3-026/2004 (program "Nanosystems, nanomaterials, and nanotechnologies", contract №38/05-N).